\documentclass[%
 reprint,
superscriptaddress,
showpacs,preprintnumbers,
 amsmath,amssymb,
]{revtex4-1}

\usepackage{graphicx}
\usepackage{dcolumn}
\usepackage{bm}
\usepackage{amsmath}
\usepackage{amssymb}

\usepackage{color,soul}
\begin{document}


\title{Polaronic Entanglement of Quantum dot Molecule in a voltage-controlled junction}%

\author{E. Afsaneh}%
 \affiliation{Department of Physics, Faculty of Science, University of Isfahan, Hezar Jerib Str., Isfahan 81746-73441, Iran.}
\author{M. Bagheri Harouni}%
\affiliation{Department of Physics, Faculty of Science, University of Isfahan, Hezar Jerib Str., Isfahan 81746-73441, Iran.}%
\affiliation{Quantum Optics Group, Department of Physics, Faculty of Science, University of Isfahan, Hezar Jerib Str., Isfahan 81746-73441, Iran}%


\begin{abstract}
We investigate the influence of vibrational phonon modes on the entanglement through a quantum dot molecule under the bias voltage-driven field. The molecular quantum dot system can be realized by coupled quantum dots in the middle of the suspended carbon nanotube. This system would be described by the Anderson-Holstein model and also can be analyzed by the polaron master equation in Markovian regime. In the presence of electron-phonon interaction, we study the entanglement as a function of bias voltage and temperature. Despite entanglement degradation because of phonon decoherence, we employ an asymmetric coupling protocol to preserve the entanglement in a significant level and also we apply the easy tunable bias voltage driven to engineer its behavior. In dynamics of entanglement, we demonstrate the phenomenon of thermal entanglement degradation and rebirth through the increase of temperature. In this process, thermal entanglement revival is intensively affected by the strength of phonon decoherence. Such that, stronger revival is occurred for higher phonon coupling amount. With an applied time-dependent bias voltage, the entanglement evolution shows periodic revival by time and in response to bias voltage rising, it illustrates decreasing and grows steadily to reach the flat form with considerable magnitude. 
\end{abstract}


%
%
\maketitle


Quantum entanglement has been considered as a crucial resource in quantum information theory which utilizes the non-local correlations without any classical analogs\cite{Peres,Horodecki,Nielsen}. Advanced technologies in quantum information processing mostly stand on single-photon controls with quantum optics applications. Implementing single electrons rather than photons utilizing quantum electron optics\cite{Hermelin,Bai} may extend researching on entanglement area by the improvement of on-chip condensed matter experiments. The entanglement of fermions as indistinguishable particles in condensed matter systems has been evaluated by fermionic concurrence\cite{Schliemann-2001,Schliemann-2001-2,Schliemann-2002,Majtey,Arxiv} in analog with Wootter's formula\cite{concurrence-1,concurrence-2}. To characterize the fermionic concurrence, quantum dot qubits containing electrons have attracted considerable attention\cite{Nori-2}. Quantum dots(QDs) as artificial atoms play prominent roles in quantum information studies where their discrete energy levels can be easily tuned by applying gate voltages\cite{Loss,Hanson-2}. 
 For realizing QDs under the influence of phonon interactions, carbon nanotube(CNT) is considered as a potential construction\cite{Postma,Graber}.  Suspended CNT which can oscillate freely is applied for preparing quantum dots\cite{Walter,Tang}. Due to the suspension of CNT,  coupled QDs which can form quantum dot molecules(QDMs) are strongly affected by the electron-phonon interaction\cite{Benyamini}. Studying the electron-phonon coupling which is influenced by surrounding harmonic lattice vibrations can determine many characteristics of quantum systems such as quantum transport \cite{Leyton, Khedri}, quantum optics\cite{Denning, Nazir-2} as well as quantum entanglement\cite{Bayer, Otten} in the presence of phonon decoherence. 
In one study about transport properties of a double quantum dot setup which is coupled to phonon modes, it was shown that under the influence of electron-phonon interaction, a negative differential conductance was given rise\cite{Walter}. 
 Also, the evaluation of the charge carrier mobility in the systems with the electron-phonon coupling was calculated by one-dimensional Holstein model in finite-temperature \cite{Prodanović}. 
Among various approaches to study the electron-phonon interaction, the coupling of quantum dot-phonon can be described with the quantum master equation in both Markovian\cite{Nazir} and non-Markovian\cite{Mogilevtsev} regimes. 
The power broadening of a quantum dot cavity-QED system which is interacting with the phonon reservoir was investigated by a time-convolutionless quantum master equation approach in the polaron frame\cite{Roy}. 
Moreover, the quantum entanglement of polaron systems were studied in two models of local and nonlocal electron-phonon coupling\cite{Stojanovic}. 
In addition, it was investigated that the quantum entanglement could be characterized for the crossover between small and large polaron regimes\cite{Zhao}. 

Relevant to quantum entanglement, in the present contribution we propose a system to provide formation of entanglement and preserve it. 
In this study, we introduce a molecular quantum dot containing the electron-phonon interaction in a biased junction as an open quantum system to investigate the effects of phonon decoherence on the quantum entanglement. 
Generally, the interaction of a quantum system with its environment is inevitable which causes quantum decoherence and in turn, leads to the loss of entanglement. This plays a major obstacle in quantum correlations. 
To reduce the influence of phonon decoherence on the dynamics of entanglement, we perform an asymmetric coupling strategy to maintain the entanglement of QDs. Also, we can engineer and control the entanglement of system by tuning the external bias voltage.
For this purpose, we apply the driven bias voltage in both types of constant and periodic time-varying fields. 
In a biased junction, the primarily well-known response of a QDM to the bias voltage is an electric transport. Here after calculating the density matrix of the system, we first obtain the current-voltage(I-V) characteristic to be sure that our setup can correctly work regarding bias voltage. 
Then, we introduce the concurrence-voltage(C-V) and concurrence-temperature(C-T) curves to investigate the polaron entanglement for our QDM system in response to an applied bias voltage. 
Through the entanglement evolution by the increase of temperature, the phenomenon of thermal entanglement degradation and rebirth are observed. Also, we can demonstrate that the phonon decoherence strength has an intensive effect on the occurrence of thermal entanglement rebirth. It exhibits that for higher phonon coupling amount, the revival entanglement magnitude is larger while for no phonon interaction, the rebirth of thermal entanglement has not happened at all. Moreover, we take into account how well the dynamics of polaron concurrence can be affected by the applied periodic voltage.
 
The outline of this paper is organized as follows. Firstly, in Theoretical Model section, we propose a model describing the quantum dot molecule with phonon modes in a bias voltage junction. Next in the Results and discussion part, we present the results for the behavior of current and concurrence when bias voltage changes are constant and time-dependent. Then, we briefly conclude in Conclusions section. Finally, in Methods part, we first derive the dynamics of  QDM setup with the phonon interaction  in the framework of the polaron master equation approach and then obtain the current and concurrence features for this system. 
\section*{Theoretical Model} \label{TheoreticalModel}
\begin{figure}[ht]
\centering
\includegraphics[scale=0.45]{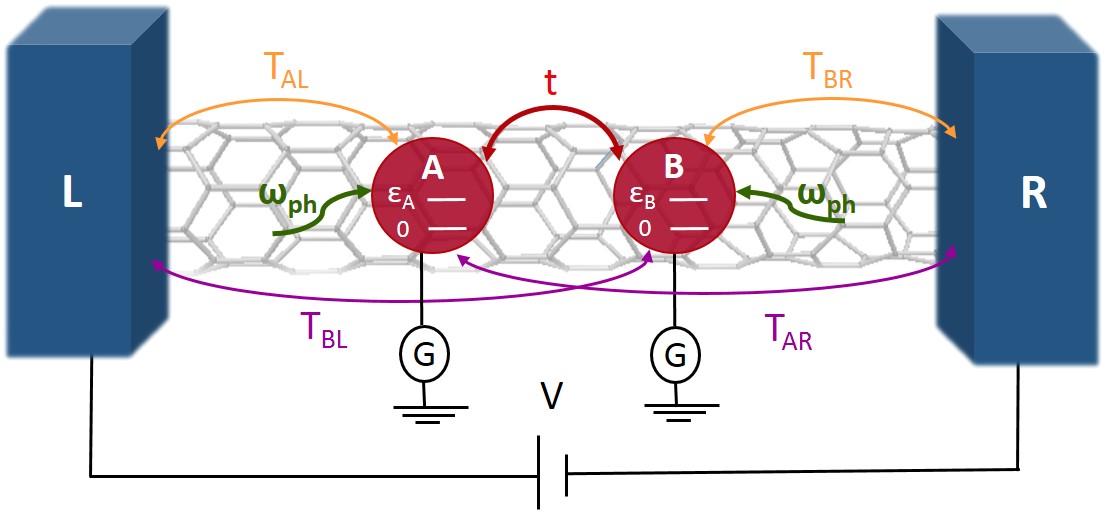}
\caption{A QDM system consists of quantum dots $A$ and $B$ are located in a suspended CNT, coupled to the local phonon modes with frequency $\omega­_{ph}$ and connected to the leads $L$ and $R$. Parameter $t$ is inter-dot coupling strength and tunnel-coupling coefficients of QDs to leads are: $T_{AL}$, $T_{AR}$, $T_{BR}$, $T_{BL}$. Energy levels of QDs, $\varepsilon_A$ and $\varepsilon_B$ are tuned by gate voltage $G$. Leads are held in potential difference $V$.}\label{Fig1}
\end{figure}

The physical system under study is a molecular double quantum dot setup in which each quantum dot is coupled to a single phonon mode in a tunnel junction schematically shown in Fig.\ref{Fig1}. 
The main purpose of this study is to investigate the quantum entanglement of the present coupled quantum dot system under the driven bias voltage.
The structure of the proposed quantum dot molecule is accomplished based on a suspended CNT. The central part of the suspended CNT is kept fixed while its lateral sides can be oscillating to perform quantum dots with electron-phonon interaction\cite{Walter,Tang}.
 Two-level quantum dots $A$ and $B$ are attached to normal metal reservoirs $L$ and $R$. These electrodes are held at chemical potentials $\mu_L$ and $\mu _R$($\mu_L$>$\mu_R$) to provide the bias voltage $V =\mu_L-\mu_R$.
The total Hamiltonian of this system is described as $H=H_{QDM}+H_{res}+H_{tun}$. To take into account the vibrational effects due to the electron-phonon interaction in the quantum dot molecule, we use the Anderson-Holstein model \cite{Holstein,Mahan}which is defined as:
\begin{eqnarray}\label{HQDM}
H_{QDM}&=&\sum_{\alpha=A,B}\left(\varepsilon_{\alpha}d^{\dagger}_{\alpha}d_{\alpha}
+\omega_{ph}b^{\dagger}_{\alpha}b_{\alpha}+t_{ph}\hat{n}_{\alpha}(b^{\dagger}_{\alpha}+b_{\alpha})\right) \nonumber \\
&+&t_{AB}(d^{\dagger}_{A}d_{B}+d_{A}d^{\dagger}_{B}).
\end{eqnarray}
Here, $H_{QDM}$ consists of Hamiltonian of quantum dots, $H_{dots}=\sum_{\alpha=A,B}\varepsilon_{\alpha}d^{\dagger}_{\alpha}d_{\alpha}+ t_{AB}(d^{\dagger}_{A}d_{B}+d_{A}d^{\dagger}_{B})$; 
Hamiltonian of phonons, $H_{ph}=\sum_{\alpha=A,B}\omega_{ph}b^{\dagger}_{\alpha}b_{\alpha}$ and Hamiltonian of electron-phonon interaction, 
$H_{el-ph}=\sum_{\alpha=A,B}t_{ph}n_{\alpha}(b^{\dagger}_{\alpha}+b_{\alpha})$. In which, $0$ and $\varepsilon_{\alpha}$ denote empty and occupied electronic energy levels of quantum dot $\alpha$,  $\omega_{ph}$ is the local phonon frequency, $t_{ph}$ shows the strength of the electron-phonon coupling and $t_{AB}$ is the inter-dot hopping amplitude which 
with no loss of generality is taken as real parameters. Moreover, $d^{\dagger}_{\alpha}(b^{\dagger}_{\alpha})$ indicates the electron(phonon) creation operator, and 
$n_{\alpha}=d^{\dagger}_{\alpha}d_{\alpha}$ is the occupation operator for quantum dot $\alpha=A,B$. It should be noted that the fermionic operator $d_{\alpha}$ 
fulfill the fermionic anti-commutator whereas $b_{\alpha}$'s is bosonic in its nature.
The Hamiltonian of reservoirs, $H_{res}=\sum_{k, \nu=L,R} \epsilon_{k,\nu} {c}^{\dagger}_{k,\nu}c_{k,\nu}$, contains the non-interacting electrons where $\epsilon_{k,\nu}$ denotes the energy level of reservoir $\nu=L,R$ and ${c}^{\dagger}_{\nu}$ creates an electron with momentum $k$ in lead $\nu$. Here, reservoirs are assumed completely polarized in which the spin of electrons can not be distinguished.
The tunneling Hamiltonian, $H_{tun}$, corresponds to tunneling between QDs and electrodes which can be described as $H_{tun}=\sum_{k,\alpha,\nu} \left(T_{\nu}c^{\dagger}_{k\nu}d_{\alpha} +H.c.\right)$. In which, $H.c.$ represents the Hermitian conjugate and $T_{\nu}$ denotes the tunnel matrix element which is assumed energy and momentum independent. 
The tunnel-coupling strength is characterized by $\Gamma_{\nu}=2\pi N^0_{\nu}|T_{\nu}|^2$. This parameter can be defined in the wide-band limit (WBL)\cite{Dong} where, both parameters $ N^0_{\nu}$, the density of states of the lead $\nu$, and $ T_{\nu}$ are assumed constant without energy dependency. This assumption allows us to have the tunnel-coupling strength as an energy-independent feature. 
Here, we consider the total tunnel-coupling strength for our QDM system as $\Gamma =\sum_{\nu}\Gamma_{\nu}$. 
Moreover, for weak tunneling to the electronic leads, the tunnel-coupling strength assumed as the smallest energy scale in this system, $\Gamma_{\nu} \ll k_{B}T$. This condition provides the lowest order of tunnel-coupling and also, it allows the leads to stay in thermal equilibrium. 

In order to proceed further, the $H_{QDM}$ in Eq.(\ref{HQDM}) is diagonalized by applying the  polaron Lang-Firsov transformation as $\bar{H}=e^{S}He^{-S}$\cite{Mahan,Lang}. In which, $S=\sum_{\alpha} g_{ph}n_{\alpha}(b^{\dagger}_{\alpha}-b_{\alpha})$ with $g_{ph}=\frac{t_{ph}}{\hbar \omega_{ph}}$ that denotes the coupling of the local phonon mode with energy $\hbar \omega_{ph}$. This transformation eliminates the electron-phonon coupling term in $H_{QDM}$(Eq.(\ref{HQDM})). So, the transformed $QDM$ Hamiltonian is obtained as:
\begin{equation}\label{TransformedHQDM}
\bar{H}_{QDM}=\sum_{\alpha}\left( \bar{\varepsilon}_{\alpha}d^{\dagger}_{\alpha}d_{\alpha}
+\hbar\omega_{ph}b^{\dagger}_{\alpha}b_{\alpha}\right) +t_{AB}\left(d^{\dagger}_{A}d_{B}+d_{A}d^{\dagger}_{B}\right),
\end{equation}
where $ \bar{\varepsilon}_{\alpha}=\varepsilon_{\alpha}-g^{2}_{ph}\omega_{ph}$  indicates the renormalized QD energy levels. As a consequence of transformation introduced in 
Eq.(\ref{TransformedHQDM}), the Hamiltonian of the uncoupled reservoirs remains unchanged while the tunneling Hamiltonian is transformed as $\bar{H}_{tun}=\sum_{k,\alpha,\nu} \left(T_{\nu}c^{\dagger}_{k\nu}X_{\alpha}d_{\alpha} + H.c.\right)$.
Here, $X_{\alpha}=e^{g_{ph}(b^{\dagger}_{\alpha}-b_{\alpha})}$ shows the polaron operator. For utilizing the prepared Hamiltonian in order to achieve the proper results for our QDM system, we define an asymmetric factor in the following. 


\subsection*{Asymmetric Factor}
In this study, employing phonon baths for our QDM setup as a dissipation process leads to entanglement degradation.
We try to compensate for the effect of phonon dissipation by applying the coupling coefficients in a left-right asymmetric way.
This may lead to keeping the amount of current and entanglement quantities in significant level.
Very recently, we defined an asymmetric factor to express the influence of unequal left and right coupling coefficients to achieve the robust entanglement for a QDM system in a Josephson junction\cite{Arxiv}. 
Here, energy-dependent parameters are the coupling coefficients of QDs with the reservoirs. These features can be taken into account left-right asymmetrically by means of tuning the gate voltages.
Usually for simplicity, only the coupling of each QD with the near-lead is considered \cite{Segal-2015}, but in this study we assume that QDs are connected to both near and far reservoirs with different non-zero couplings.  Therefore, we deal with the tunnel-coupling strength for each reservoir as $T_{\nu}=\frac{T_{A\nu}+T_{B\nu}}{2} $. In which $T_{A\nu}(T_{B\nu})$ is the coupling strength of $QD_{A}(QD_{B})$ with the reservoir $\nu$. In other words, our system needs four tunnel-coupling coefficients of $ T_{AL}$, $ T_{BL}$, $ T_{AR}$ and $ T_{BR}$ which are shown in Fig.\ref{Fig1}. 
The asymmetry of coupling strengths can be demonstrated in the asymmetric factor definition $\kappa=\frac{\kappa_{A}+\kappa_{B}}{2}$ where $\kappa_{\alpha}=|\frac{T_{\alpha L}-T_{\alpha R}}{T_{\alpha L}+T_{\alpha R}}|(\alpha=A, B)$\cite{Arxiv}. 
The amount of asymmetric factor ranges from zero for symmetric coupling coefficients, to unit magnitude for completely asymmetric ones.

The symmetric structure is defined when each QD is coupled to the left and right leads with the same coupling coefficient, $T_{\alpha L}=T_{\alpha R}$. This situation provides minimum magnitude of asymmetric factor, $\kappa=0$. The asymmetric structure is referred to the left-right different coupling coefficients, $T_{\alpha L} \neq T_{\alpha R}$, with $0<\kappa\leq1$ amount. Completely asymmetric configuration with $\kappa \simeq 1$ can be realized for the specific physical properties when the strength coupling of near-lead is much larger than the far-lead. 
This means that we have $T_{AL} \gg T_{AR}$($T_{BR} \gg T_{BL}$) for $QD_A$($QD_B$). In the next section, we present the obtained results for the present QDM system and discuss about them.
\section*{Results and discussion}\label{Result}
In this section, we present the results of investigating both polaron electric current and polaron concurrence behaviors in response to the bias voltage as well as temperature. Moreover, the dynamics of the system for two situations of applying constant and harmonic time-varying bias voltage are studied.
For simplicity and with no loss of generality, the characteristics of local phonon near each QD $\alpha=A, B$ are assumed  the same. Therefore, phonon resonances are defined similar to each other $\omega_{A}=\omega_{B}=\omega_{ph}$ and also, the phonon temperatures $ T_A$ and $ T_B$ are selected  the same and as equal as the temperature of electrons in reservoirs, $T_A=T_B=T_{el}$. 
\subsection*{Constant Voltage}
In this part, we suppose that the QDM is driven by a dc bias voltage. First, we calculate the polaron electric current as a function of bias voltage and present the I-V 
characteristic. Then we investigate the dependence of the concurrence on the constant bias voltage as well as temperature through the studying C-V and C-T characteristic curves.
\subsubsection*{Current-Voltage}
\begin{figure}[ht]
\centering
\includegraphics[scale=0.23]{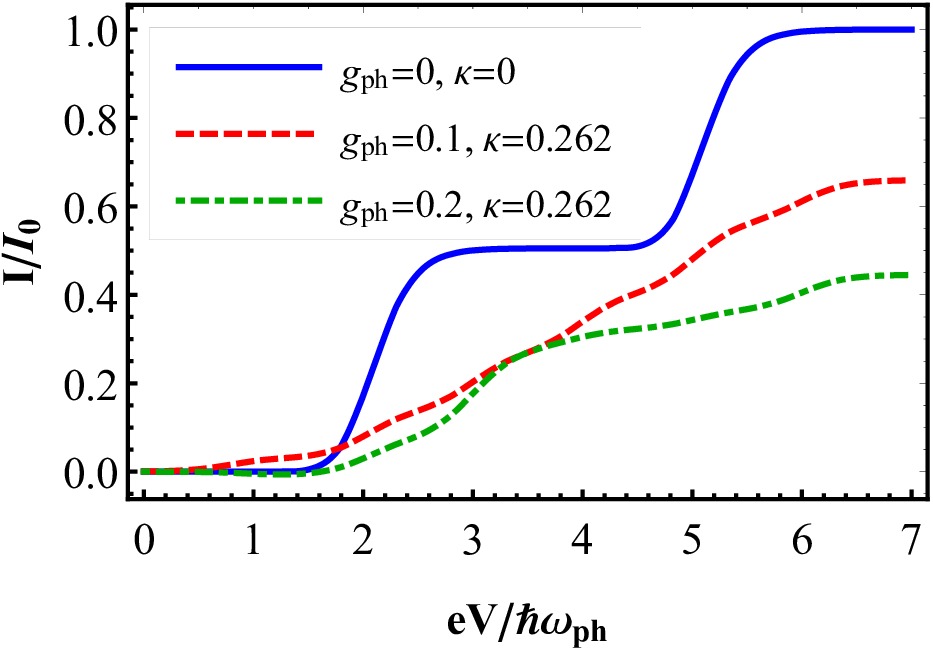}
\caption{Current-Voltage behavior in the presence of phonon in quantum dots for $\frac{k_{B}T}{\hbar\omega_{ph}}=0.2$ and  $I_0=e\frac{\Gamma_0}{\hbar}$.}\label{Fig2}
\end{figure}

In Fig.\ref{Fig2}, we show I-V curve through the molecular double dot system for different magnitudes of electron-phonon coupling, $g_{ph}$.
In the absence of phonons, $g_{ph}=0$, we consider the symmetric coupling coefficient, $\kappa=0$, while for non-zero electron-phonon coupling, QDs are prepared asymmetrically with $\kappa=0.262$.
In this figure, we see that in a certain electron-phonon coupling, the behavior of current is normally growing due to the increase of bias voltage. Moreover, in a determined bias voltage, the current has lower magnitude for the higher amount of electron-phonon coupling coefficients.
In the case of $g_{ph}=0$, when energies are in resonant with the energy level of QDs which assumed $\frac{\varepsilon_{A}}{\hbar\omega_{ph}}=2$ and $\frac{\varepsilon_{A}}{\hbar\omega_{ph}}=5$, the current illustrates steps and for high bias voltage, it receives to a flat form. In the presence of electron-phonon coupling for $g_{ph}=0.1$ and $g_{ph}=0.2$, the amount of current is reduced. Also, the current curves exhibit smoother such that the step structure are completely disappeared.
\subsubsection*{Concurrence-Voltage}
\begin{figure}[t]
\begin{minipage}[t]{0.9\linewidth}
\centering
\includegraphics[width=0.9\textwidth]{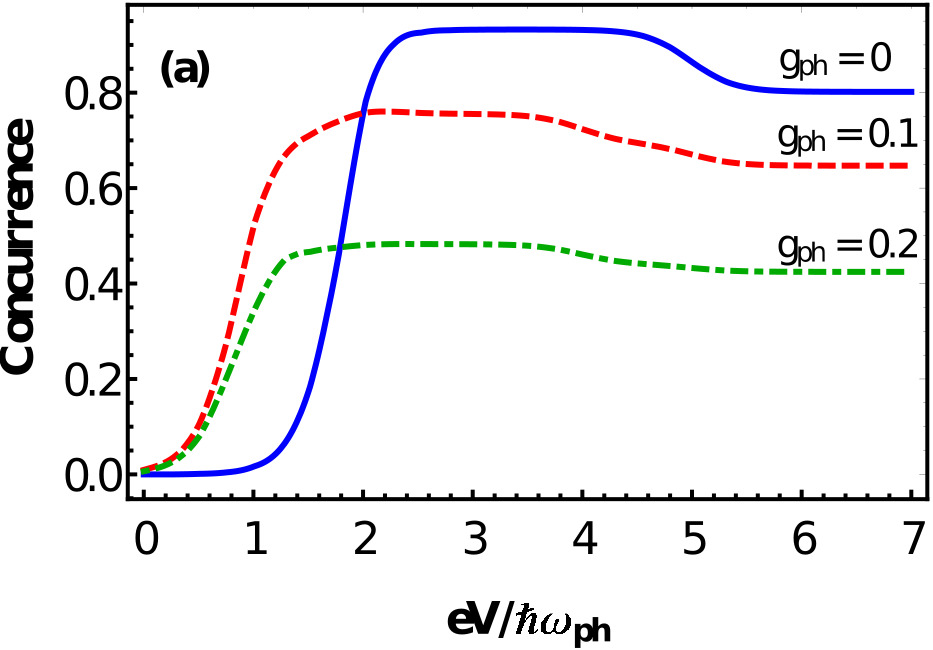}
\end{minipage}
\hspace{0.1cm}
\begin{minipage}[t]{0.9\linewidth}
\centering
\includegraphics[width=0.9\textwidth]{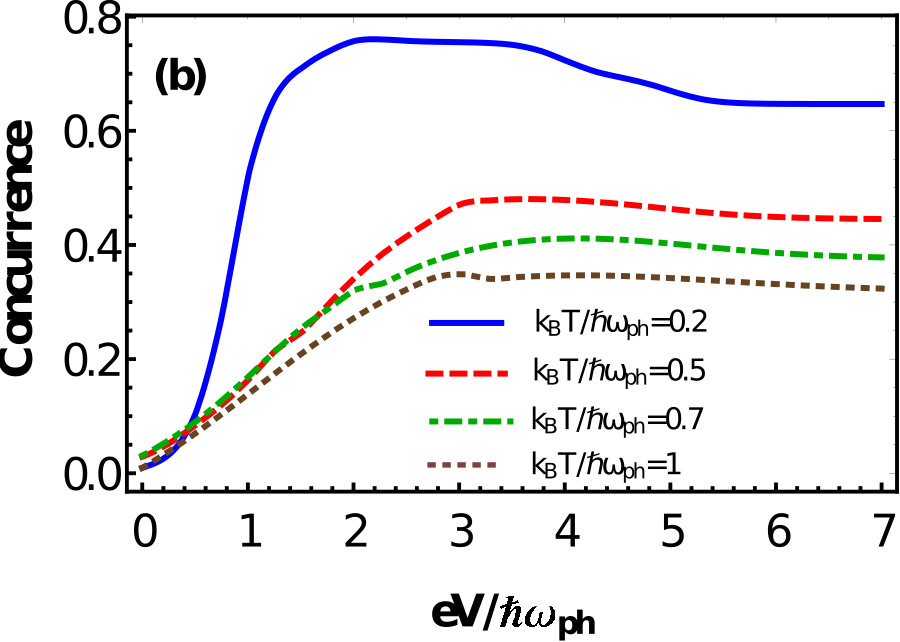}
\end{minipage}
\caption{Concurrence-Voltage behavior for $\kappa=0.55$, (a): in certain electron-phonon coupling amounts with $\frac{k_{B}T}{\hbar\omega_{ph}}=0.2$. (b): in certain temperatures with $g_{ph}=0.1$.}\label{Fig3}
\end{figure}

The concurrence dependency on the bias voltage is illustrated in C-V characteristic curve, Fig.\ref{Fig3}, with two panels for asymmetric coupling, $\kappa=0.55$.
Panel $a$ of this figure demonstrates the influence of electron-phonon coupling strength on the concurrence. 
In this figure when there is no phonon coupling $g_{ph}=0$, the concurrence is increased by the increment of bias voltage which is similar the behavior of I-V curve(Fig.\ref{Fig2}). 
Also in resonant points of QDs energies, the concurrence exhibits steps. Moreover, in high bias voltage, it reaches to the steady-state. 
The concurrence curves for non-zero coupling(dashed and dot-dashed lines) follow the behavior of no phonon curve(solid line) only with lower magnitude and a little smoother. 
In panel (b) of Fig.\ref{Fig3}, we plot C-V curve for a fixed phonon coupling in different temperatures. This figure shows that the magnitude of concurrence is decreased at higher temperatures. 
Generally in Fig.\ref{Fig3}, both parameters of increasing the electron-phonon coupling strength in panel $a$ and rising the temperature in panel $b$ lead to the entanglement degradation. Despite these two dissipative features, the concurrence amount is kept in significant level. The main reason of this behavior of system for preserving the entanglement is due to applying the coupling coefficients asymmetrically with non-zero asymmetric factor. 
\subsubsection*{Concurrence-Temperature}
\begin{figure}[ht]
\centering
\includegraphics[scale=0.23]{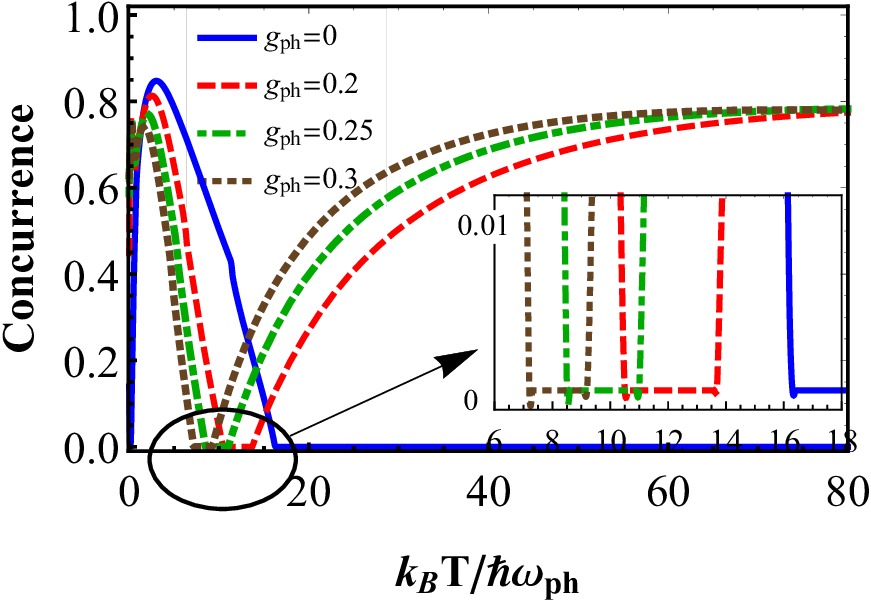}
\caption{Concurrence-Temperature behavior for $\frac{eV}{\hbar\omega_{ph}}=0.1$ and $\kappa=0.63$.}\label{Fig4}
\end{figure}

To obtain more information about the evolution of entanglement in the present system, the dependence of concurrence on the temperature(C-T characteristic) is shown in Fig.\ref{Fig4}.
In this figure for an applied constant bias voltage, unentangled electrons can move from the initial ground states to the excited ones. Indeed, for the higher amount of temperature, electrons which are scattered by phonons repeatedly can find more opportunities for excitation which means that through their motivation they can have more possibilities to be entangled.   
This behavior of electrons provides an increase of concurrence up to receive the most magnitude in the relevant condition. 
 Rising more temperature leads to the separation of the entangled electrons which can be observed by the concurrence decrease manner. 
This decline trend which is related to the loss of coherence continues through a finite temperature interval until it reaches the zero amount when all electrons being unentangled. Then, for more temperature, entanglement appears again. We describe the whole behavior of concurrence-temperature characteristics with two definitions, Thermal entanglement degradation and Thermal entanglement rebirth as follows. \\ 

\begin{itemize}
\item \textbf{Thermal entanglement degradation}
We call the complete disappearance of entanglement with respect to the increase of temperature in the present system as thermal entanglement degradation(TED). 
In Fig.\ref{Fig4}, TED process exhibits that for stronger decoherence strength which is due to the larger phonon coupling can cause shorter temperature interval and also smaller magnitude of concurrence maximum. 
One of the most remarkable features of this model can be observed by applying more temperature to obtain thermal entanglement rebirth. \\  

\item \textbf{Thermal entanglement rebirth}
In C-T curves of Fig.\ref{Fig4} with electron-phonon coupling(dashed, dot-dashed and dotted lines), the entangled electrons which are separated completely after the thermal entanglement collapse are occasioned to be entangled again in response to more temperature. This causes the reappearance of concurrence after its complete disappearance. 
We name the reappearing of the entanglement regards to temperature as thermal entanglement rebirth(TER). 
\end{itemize}

Interestingly, we discover that for stronger phonon coupling, concurrence reappears more sever in TER phenomenon which is shown in Fig.\ref{Fig4}. In other words, the reappearance of thermal entanglement with $g_{ph}=0.3$(dotted line) revives to larger magnitude than the curves with $g_{ph}=0.25$ and $g_{ph}=0.2$(dot-dashed and dashed lines respectively). However, for the curve with no phonon coupling $g_{ph}=0$(solid line), there is no rebirth occurrence for the thermal entanglement. This means that when QDM system is determined as an open quantum system coupled with only thermal baths, exclusively the thermal entanglement degradation occurres without any thermal entanglement revival.
This phenomenon represents that the occurrence of concurrence rebirth is due to the decoherence of surrounding phonons which is thoroughly controlled the amplitude of revivals. 
Furthermore, we find out that concurrence continues in a steady way to reach the flat form with significant level regards to the increase of temperature. The main point of this treatment is that despite the decoherence of phonon as well as the dissipation of rising temperature, the entanglement of the proposed setup is kept robustly. Obviously, this robustness shows the capability of the present model for preserving the entanglement. 
However, usually the revivals of quantum correlations have occurred in the multipartite systems\cite{Lopez, Cakmak}, in qubits system affected by environments such as the non-Markovian reservoirs\cite{Mazzola} and the classical environments\cite{Xu} or in other complicated structures.  In this study which is related to a very simple bipartite setup surrounded by phonon bath, we can observe not only the revival entanglement but also protect the entanglement steadily through the increase of temperature.
\subsection*{Time-dependent Voltage}
To further investigate the parameters which affect the entanglement of QDM setup corresponding to the phonon decoherence, we apply time-dependent voltage. 
Usually, the time evolution of entanglement under the periodic driven was studied in structures with multi subsystems\cite{Kim, Ponte}. While, we try to investigate the entanglement when is driven periodically in both cases of through the evolution of time and in response to external bias voltage for our bipartite system. 
Here, we implement an ac periodic voltage to the quantum dot molecule as: 
\begin{equation}
V(t)=V^{0}_{dc}+V^{0}_{ac}Cos(\omega_{ac}t),
\end{equation}
where $V^{0}_{dc}$ indicates the amplitude of constant voltage while $V^{0}_{ac}$ and $\omega_{ac}$ denote the amplitude and frequency of oscillating voltage, respectively.
By applying time-dependent voltage to the system, some parameters can change depending on time. These elements are energy level of QDs, electrochemical potential of reservoirs and tunnel-coupling strength. As, we suppose our system in WBL regime, therefore the tunnel-coupling coefficients are assumed constant and energy-independent parameters. In the following first, we focus on the response of current and then polaronic entanglement on the periodic voltage.
\subsubsection*{Current}
\begin{figure}[t]
\begin{minipage}[t]{0.9\linewidth}
\centering
\includegraphics[width=0.9\textwidth]{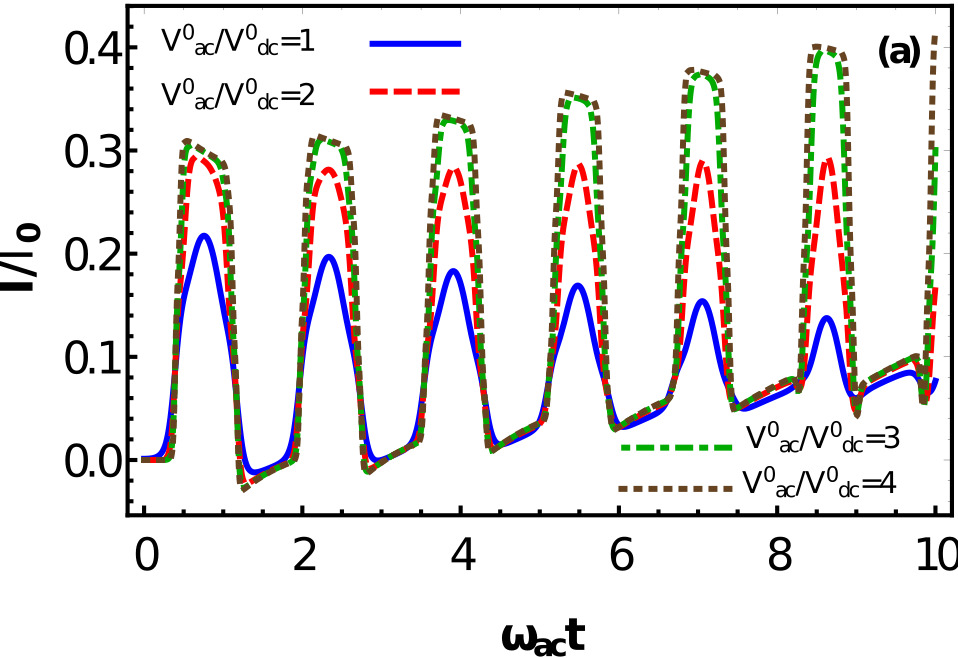}
\end{minipage}
\hspace{0.1cm}
\begin{minipage}[t]{0.9\linewidth}
\centering
\includegraphics[width=0.9\textwidth]{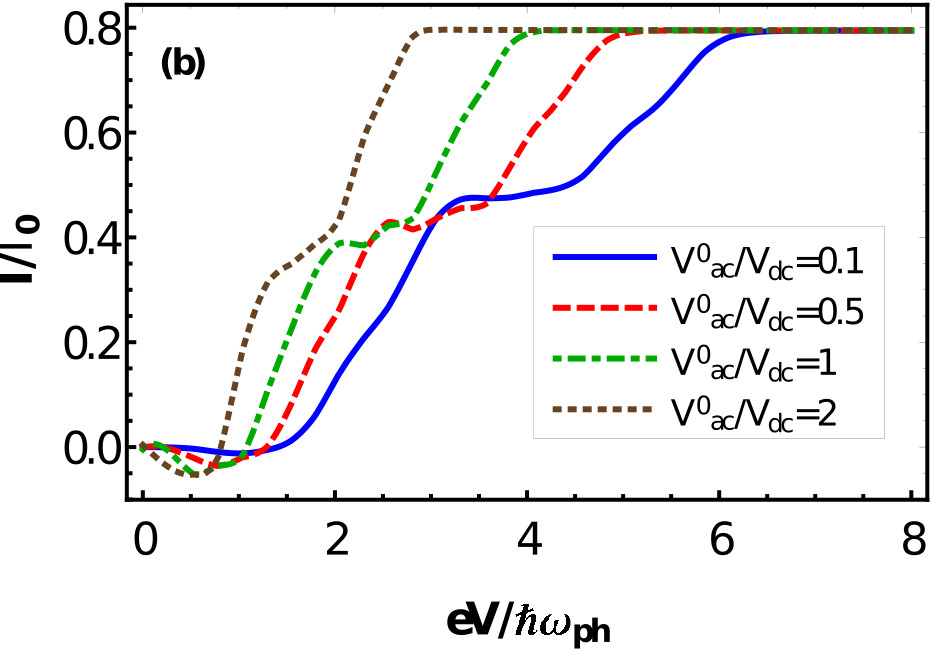}
\end{minipage}
\caption{Current evolution for periodic voltage $V_{ac}=V_{ac}^0 Cos(\omega_{ac}t) $ with $\frac{eV^{0}_{dc}}{\hbar \omega_{ph}}=1$, $g_{ph}=0.1$, $\kappa=0.58$ and $\frac{k_{B}T}{\hbar \omega_{ph}}=0.2$ (a): Current time evolution curve with $\frac{\omega_{ac}}{\omega_{ph}}=4$ and (b): Current-voltage characteristic with $\frac{\omega_{ac}}{\omega_{ph}}=0.5$. Here, $I_0=e\frac{\Gamma_0}{\hbar}$.}\label{Fig5}
\end{figure}

Figure \ref{Fig5} illustrates the dynamics of current under the harmonic voltage for a fixed phonon coupling, $g_{ph}=0.1$. In panel (a), the varying time current can oscillate stronger for higher amounts of amplitude with the frequency modulation $\frac{\omega_{ac}}{\omega_{ph}}=4$. 
Panel (b) exhibits the current-voltage characteristic in which current increases more quickly and receives to steady state faster for higher amplitude amounts in low frequency modulation $\frac{\omega_{ac}}{\omega_{ph}}=0.5$ of periodic voltage. 

\subsubsection*{Concurrence}
\begin{figure}[t]
\begin{minipage}[t]{0.9\linewidth}
\centering
\includegraphics[width=0.9\textwidth]{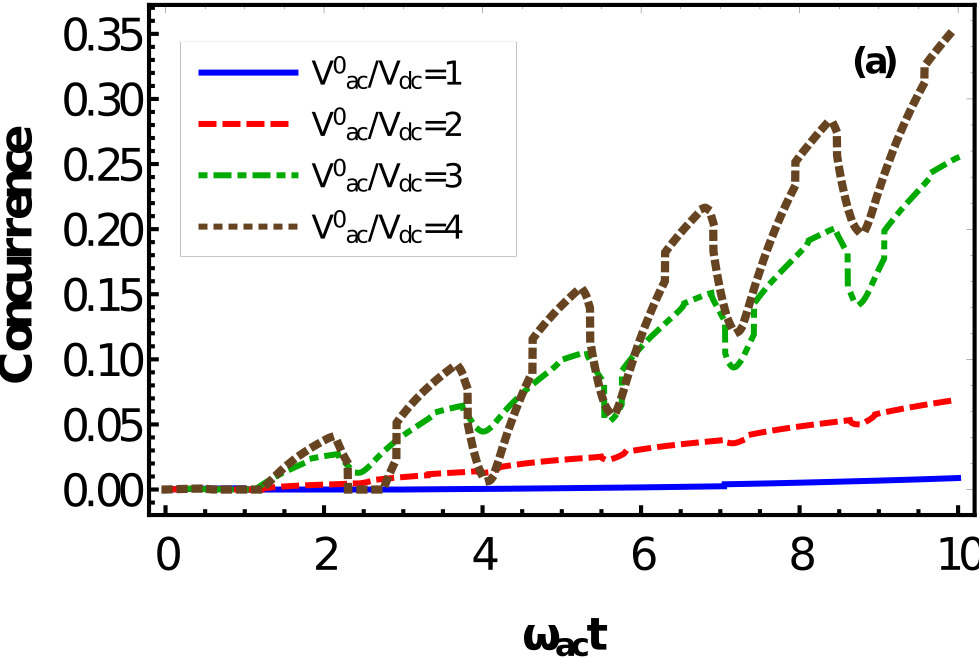}
\end{minipage}
\hspace{0.1cm}
\begin{minipage}[t]{0.9\linewidth}
\centering
\includegraphics[width=0.9\textwidth]{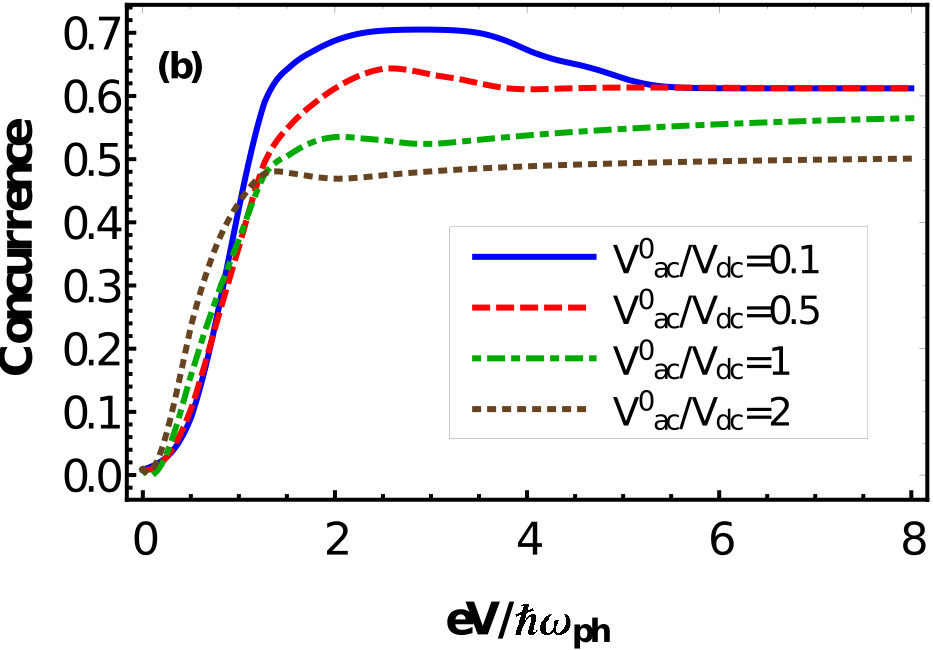}
\end{minipage}
\caption{Concurrence evolution for periodic voltage $V(t)=V_{dc}^0+V_{ac}^0 Cos(\omega_{ac}t) $ with $\frac{eV^{0}_{dc}}{\hbar \omega_{ph}}=1$, $g_{ph}=0.1$, $\kappa=0.58$ and $\frac{k_{B}T}{\hbar \omega_{ph}}=0.2$ (a): Concurrence time evolution with $\frac{\omega_{ac}}{\omega_{ph}}=4$ and (b): Concurrence-voltage characteristic with $\frac{\omega_{ac}}{\omega_{ph}}=0.5$}\label{Fig6}
\end{figure}

The dynamics of concurrence is shown in Fig.\ref{Fig6}.  
Panel (a) of this figure demonstrates that time evolution of concurrence can revive periodically.  
In each driving cycle, the oscillating concurrence gradually increases to receive the maximum amount and then it decreases to reach a point with higher magnitude than the starting point of cycle. In other words, after a complete cycle, it begins to revive with bigger entanglement amount. In curves with larger amplitudes, concurrence grows quickly to reach the higher values. Comparing this figure with Fig.\ref{Fig5}a shows that both of concurrence and current are rising through the time. Only, the dynamics of current changes by time severely oscillating than the concurrence.

In contrast with concurrence time evolution(Fig.\ref{Fig6}a), C-V characteristic of time-dependent voltage(Fig.\ref{Fig6}b) demonstrates lower amount for larger amplitudes in certain oscillating frequency and fixed electron-phonon coupling. However, panel (b) of Fig.\ref{Fig6} illustrates decreasing for a given amplitude, it follows steadily to achieve a flat form with a considerable amount. This C-V curve shows that the entangled electrons never disentangled by the time-dependent driving signal. This behavior demonstrates the strength of the introduced QDM setup to protect the entanglement of electrons under the time-varying bias voltage control in spite of the phonon decoherence. Drawing a comparison between I-V curve(Fig.\ref{Fig5}b) with C-V characteristic(Fig.\ref{Fig6}b), one finds that for larger amount of amplitude, the electric current finds higher magnitude while the polaron concurrence behaves reversely.  This opposite response may originate from the behavior of electrons with respect to the time-dependent bias voltage. By increasing voltage, electrons can transfer more quickly which leads to larger current amount. However when electrons move oscillating faster, the probability of electron presence which can be entangled is decreased that causes to diminishing the entanglement amount. 
\section{Conclusions}\label{Conclusion}
We investigated the polaronic entanglement in a molecular quantum dot system through the concurrence. The influence of bias voltage and temperature on the concurrence were studied in the presence of electron-phonon coupling.  Since, the phonon decoherence caused the entanglement degradation, we implemented the coupling of QDs asymmetrically to preserve the entanglement. Also we applied the constant as well as time-dependent bias voltage to control the entanglement magnitude. In concurrence-temperature curve, the thermal entanglement degradation and rebirth occurred regarding the increase of temperature. In this phenomenon, the strength of phonon coupling affects the rebirth of thermal entanglement comprehensively. In which, for higher phonon coupling amount,  revival is occurred stronger.  For external time-dependent bias voltage, QDM system demonstrated periodic revival. In this system,  concurrence time evolution showed rising in certain amplitudes while for entanglement in response to voltage increase, it received a plat form.  
\section{Methods}\label{Method}
\subsection{Dynamics}\label{Dynamics}
To explore the dynamics of the present molecular quantum dot system in the presence of electron-phonon coupling and to obtain both quantum transport and quantum entanglement, we start from the Liouville-von Neumann equation of the whole system in the interaction picture\cite{Breuer}.  
The complete system consists of the central double quantum dot, electronic reservoirs and oscillating phonon baths systems with Hilbert spaces $\mathcal{H}_{dots}$, $\mathcal{H}_{res}$ and $\mathcal{H}_{ph}$ respectively. 
Consequently, the total Hilbert space of the whole system is defined as $\mathcal{H}_{tot}=\mathcal{H}_{dots}\otimes \mathcal{H}_{res}\otimes \mathcal{H}_{ph}$.
We suppose that the present system is considered in high enough phonon frequency to obtain the strong electron-phonon coupling.  Also, for assumption of weakly coupling of quantum dots with reservoirs, we can apply Born-Markov approximation.
Therefore, the density matrix of the total system is approximately characterized by $\rho_{tot}(t) \approx \rho_{dots}(t)\otimes\rho_{res}(t)\otimes\rho_{ph}(t)$ for 
thermal equilibrium baths. By tracing out the bath degrees of freedom for both reservoirs and phonon baths, the quantum master equation of central QDM system is obtained as:
\begin{eqnarray}\label{QME}
\frac{d\rho_{s}(t)}{dt}&=&-\frac{i}{\hbar} [\bar{H}_{dots}(t),\rho_{s}(t)] \nonumber \\
&-& \frac{1}{\hbar^2}\int_{0}^{\infty}dt^{'}tr_{res}tr_{ph}\left[ \bar{H}_{tun}(t)\left[\bar{H}_{tun}(t-t^{'}),\rho_{tot}(t)\right] \right].\nonumber \\
\end{eqnarray}
The first term corresponds to a coherent evolution of system while the second one arises due to the different sources of dissipation. We ignore the first term in the condition of  $t_{AB}\leq T_{\nu}\leq t_{ph}$ which means that electron-phonon coupling is larger than the coupling of dots with reservoirs, so the coherent dynamics of dots is negligible in the presence  of phonon interaction.
 By substituting the transformed Hamiltonians, $\bar{H}_{QDM}$ and $\bar{H}_{tun}$ into Eq.(\ref{QME}), the reduced density matrix fulfills the following equation:
\begin{eqnarray}
\frac{d\rho_{s}}{dt}&=&\sum_{\alpha,\nu} (M^{+}_{\nu}\left[2d_{\alpha}\rho_{s}(t)d^{\dagger}_{\alpha}-d^{\dagger}_{\alpha}d_{\alpha}\rho_{s}(t)-\rho_{s}(t)d^{\dagger}_{\alpha}d_{\alpha} \right]\nonumber \\
&+&M^{-}_{\nu}\left[ 2d^{\dagger}_{\alpha}\rho_{s}(t)d_{\alpha}-d_{\alpha}d^{\dagger}_{\alpha}\rho_{s}(t)-\rho_{s}(t)d_{\alpha}d^{\dagger}_{\alpha} \right]).
\end{eqnarray}
In this equation, $M^{\pm}_{\nu}$ describe the tunneling rates as:
\begin{eqnarray}\label{M+}
M^{+}_{\nu}&=&\sum_{\alpha,\nu} \Gamma_{\nu} \int d\omega \langle c_{\nu}c^{\dagger}_{\nu} \rangle \langle X^{\dagger}_{\alpha}X_{\alpha}\rangle \nonumber \\
&=&\Gamma_{\nu} \int d\omega [1-f_{\nu}(\omega+\epsilon_{\nu})]G^{+}_{\alpha}(\omega),
\end{eqnarray}
and
\begin{eqnarray}\label{M-}
M^{-}_{\nu}&=&\sum_{\alpha,\nu}\Gamma_{\nu} \int d\omega \langle c^{\dagger}_{\nu}c_{\nu} \rangle \langle X_{\alpha}X^{\dagger}_{\alpha}\rangle \nonumber \\
&=&\Gamma_{\nu} \int d\omega f_{\nu}(\omega+\epsilon_{\beta})G^{-}_{\alpha}(\omega).
\end{eqnarray}
In Eq.(\ref{M+}), the correlation function of reservoirs is defined as $\langle c_{\nu}c^{\dagger}_{\nu} \rangle=tr_{res}[\rho_{res}c_{\nu}c^{\dagger}_{\nu}]$  which gives us the Fermi distribution function of lead $\nu$ by $f_{\nu}(\omega+\epsilon_{\nu})=\frac{1}{e^{\beta_{el}(\omega+\epsilon_{\nu})}+1}$. Temperature of electrons in both leads is assumed the same and is shown by $T_{el}$ which provides $\beta_{el}=\frac{1}{k_BT_{el}}$.
The polaron correlation functions are demonstrated by $G^{\pm}_{\alpha}(\omega)$. Here, negative correlation function is given by $G^{-}_{\alpha}(t)=tr_{ph}[\rho_{ph}X_{\alpha}(t)X^{\dagger}_{\alpha}]$\cite{Mahan}, and its Fourier transform which is $G^{-}_{\alpha}(\omega)=\int^{\infty}_{0} dt e^{i \omega t}G^{-}_{\alpha}(t)$ can be calculated as:
\begin{eqnarray}
G^{-}_{\alpha}(\omega)&=&e^{-g(2N_{ph}+1)}\sum^{\infty}_{l=-\infty}I_{l} \{ 2g\sqrt{N_{B}(1+N_{B})} \} \nonumber \\
&\times & e^{l\frac{\omega_{ph}}{2}\beta}2\pi \delta(\omega-\omega_{ph}l).
\end{eqnarray}
Here, Bose-Einstein distribution function of phonon is $N_{ph}=\frac{1}{e^{\beta_{ph}\omega_{ph}}-1}$ with $\beta_{ph}=\frac{1}{k_BT_{ph}}$. Since phonons inside the CNT are close to leads, phonon temperature is assumed equal to the electron temperature of leads, $T_{ph}=T_{el}$. $I_l$ denotes the modified Bessel function of the first kind of order $l$. The positive and negative correlation functions are related to each other by the relation $G^{+}_{\alpha}(\omega)=G^{-}_{\alpha}(-\omega)$.
Now with the calculated polaron correlation function of system, we can proceed further to obtain the quantities of polaron transport as well as polaron concurrence. 
In the following, current ang concurrence parameters for the present system are calculated. 
\subsection{Current}\label{Current}
To have quantum transport through the QDM system, an external bias voltage $V$ is applied across the junction from the left reservoir which is illustrated in Fig.\ref{Fig1}. Therefore, electric current flows from the left lead as a source into the dots and follows to the right lead as a drain.
Electric current out of the lead $\nu $ can be defined as the change of the expectation value of the particle numbers\cite{Mahan}:
\begin{eqnarray} \label{current}
I_{\nu }(t)&=& -e \frac{d}{dt}\langle N \rangle_{\nu}=\frac{ie}{\hbar}[N_{\nu}(t),H_{tun}(t)]\nonumber \\
&=& \frac{ie}{\hbar} \sum_{k\nu} (T_{\nu} \hat{c}^{\dagger}_{k \nu} \hat{d}_{\alpha} - H.c.),
\end{eqnarray}
in which $N_{\nu}=\sum_{\nu} c^{\dagger}_{\nu} c_{\nu} $.
The QME relation in Eq.(\ref{QME}) can be written again in the matrix form as $\dot{\rho}=M\rho$, where the matrix $M$ corresponds to the characteristics of  the master equation. According to the matrix formalism of QME, we can express the electric current as \cite{Mukamel,Afsaneh, Arxiv}:
\begin{eqnarray} \label{current}
\hat{I}_{\nu }(t)&=& -e \frac{d}{dt} \langle N \rangle_{\nu }=-e \frac{d}{dt} tr_{\nu }[N_{\nu } \rho], \nonumber \\
&=& \frac{e}{\hbar} \langle N| M_{\nu } | \rho \rangle
\end{eqnarray}
where $ M_{\nu }$ denotes the contribution of lead $\nu$ in matrix $M$.
Here, we study the stationary transport which is obtained when the system takes long enough time($t \rightarrow \infty$) to reach the steady state.   
\subsection{Concurrence}\label{Concurrence}
To study the entanglement of the quantum dot systems, concurrence is an appropriate measure.
The entanglement of QDs which consists of electrons as indistinguishable particles should be calculated in the formalism of fermionic concurrence.
In the previous paper\cite{Arxiv}, we discuss that the fermionic concurrence for the indistinguishable particles can be characterized by alogue with the Wootters’ formula.
To quantify the entanglement of two qubits such as quantum dots,  Wootters proposed the measure of concurrence for both pure and mixed states\cite{ concurrence-2}. 
The concurrence of two QDs is defined as:
\begin{equation}\label{concurrence-eq1}
C(\rho)=Max [0,\lambda_1-\lambda_2-\lambda_3-\lambda_4],
\end{equation}
where, $\lambda_i,(i=1,2,3,4)$ are the non-negative eigenvalues of matrix $R$ in decreasing order, $\lambda_1>\lambda_2>\lambda_3>\lambda_4$. Matrix $R$ is evaluated as $R=\sqrt{\sqrt{\rho}\tilde{\rho}\sqrt{\rho}}$ with the density matrix of the system, $\rho $, and $\tilde{\rho}=(\sigma_y \otimes \sigma_y) \rho ^*(\sigma_y \otimes \sigma_y)$.  Here, $\rho^*$ shows the complex conjugate of the density matrix  and $\sigma_y$ represents the $y$ element of Pauli matrices.
The concurrence takes value in the interval between zero, for the separable state, and one unit magnitude for the maximally entangled states.

The basis states of the present QDM system with quantum dots $A$ and $B$(Fig.\ref{Fig1}) can be selected as $|\psi\rangle_{AB}=|\Phi\rangle_{A} \otimes |\Phi\rangle_{B}$. In which, $|\Phi\rangle_{A}$ and $|\Phi\rangle_{B}$ denote the states of quantum dot $A$ and $B$, respectively. We suppose each dot with two energy levels containing unoccupied $|0\rangle_{\alpha} $ and occupied $|1\rangle_{\alpha} $ states with the energies $0$ and $\varepsilon_{\alpha}$($\alpha=A, B$), respectively. Therefore, we can present the total form of the occupation states as $|0_{A},1_{A},0_{B},1_{B}\rangle=|0_{A},1_{A}\rangle\otimes |0_{B},1_{B}\rangle $.
Here, to obtain the polaron concurrence of the QDM system, we assume that QDs are not entangled initially. This means that the environment of the system starts from the state of $|0,0,0,0\rangle$. Thus, we consider the initial state of two coupled unentangled QDs as:
\begin{equation}\label{initial-asymmetric}
\rho(0)= \left[ {\begin{array}{cccc}
   1 & 0 & 0 & 0 \\
    0 & 0 & 0 & 0  \\
    0 & 0 & 0 & 0 \\
    0 & 0 & 0 & 0 \\
  \end{array} } \right].
\end{equation}
To reach the maximum amount of entanglement for QDM with an unentangled initial state in a biased junction, we consider the coupling coefficients between the system-leads asymmetrically similar the previous research\cite{Arxiv}.


\end{document}